\documentclass[twocolumn,english,superscriptaddress,nofootinbib,preprintnumbers, aps]{revtex4-1}
\usepackage{babel}
\usepackage{amsmath}
\usepackage{graphicx}
\usepackage{amssymb}
\usepackage{ulem}
\usepackage{soul}
\usepackage{float}
\usepackage{subfig}
\usepackage{hyperref}
\usepackage{enumerate}

\usepackage{xcolor}
\definecolor{cblue}{RGB}{16,78,139}
\definecolor{cred}{RGB}{139,37,0}
\definecolor{cgreen}{RGB}{0,139,0}

\usepackage[left=2cm, right=2cm, top=2cm, bottom=2cm]{geometry}


\newtheorem{conjecture}{Conjecture}

\newtheorem{observation}{Observation}

\newcommand{\appropto}{\mathrel{\vcenter{
  \offinterlineskip\halign{\hfil$##$\cr
    \propto\cr\noalign{\kern2pt}\sim\cr\noalign{\kern-2pt}}}}}

\newcommand{\id}{\mathbb{I}}
\definecolor{lasallegreen}{rgb}{0.03, 0.47, 0.19}

\begin{document}

\title{Challenges in recovering a consistent cosmology from the effective
dynamics of loop quantum gravity}

\author{Andrea Dapor}
\email{adapor1@lsu.edu}
\affiliation{Department of Physics and Astronomy, Louisiana State University, Baton Rouge, LA 70803, USA}

\author{Klaus Liegener}
\email{liegener1@lsu.edu}
\affiliation{Department of Physics and Astronomy, Louisiana State University, Baton Rouge, LA 70803, USA}

\author{Tomasz Paw{\l}owski}
	\email{tomasz.pawlowski@uwr.edu.pl}
	\affiliation{Institute for Theoretical Physics, Faculty of Physics and Astronomy, University of Wroc{\l}aw, pl. M. Borna 9, 50-204  Wroc{\l}aw, Poland}


\begin{abstract}
We reexamine a set of existing procedures aimed at recovering the effective description of the dynamics of LQG in the context of cosmological solutions.
In particular, the studies of those methods, to which the choice of cuboidal graphs and graph-preserving Hamiltonian is central, result in the formulation of a set of no-go statements, severely limiting the possibility of recovering a physically consistent effective dynamics this way.  
% 
% 
% This concise note aims at addressing various speculations about ``effective dynamics'' that are as of today in circulation in the Loop Quantum community. The conjecture of ``effective dynamics'' states that the expectation value of the Hamiltonian in suitable families of coherent states can be used as a generator of the effective trajectory of the mean of said states. While this conjecture has been verified in the setting of Loop Quantum Cosmology, its validity in the full theory remains unproven.\\
% We investigate whether this conjecture can hold true for the so-called $\bar{\mu}$-regularisation scheme of the Hamiltonian constraint and find the answer in the negative.
\end{abstract}

\maketitle
\section{Introduction}
Past work in the area of Loop Quantum Cosmology (LQC) \cite{Bojowald:2008zzb,Ashtekar:2011ni,Bojowald:2001xe} allowed one to probe the dynamics of homogeneous cosmological systems on the genuine quantum level. A major result of these studies was the finding that the the Big Bang singularity is replaced by a bounce \cite{APS06a}. A further remarkable outcome was that the quantum trajectories are reproduced by a simple phenomenological model constructed by replacing the fundamental LQC operators with their expectation values (implicitly evaluated on certain semiclassical states). This framework is known in the literature as the {\it effective dynamics} \cite{Singh:2005xg}.

Since LQC is an independent theory never derived from Loop Quantum Gravity (LQG) \cite{Thiemann:2007zz,Ashtekar:2004eh}, the question whether the full theory would lead to similar dynamical predictions is highly nontrivial. The direct computation of the genuine quantum dynamics in LQG is outside of technical reach (except in some unphysical toy examples \cite{AIL17,Zhang:2019dgi}). Observation of the success of effective dynamics in LQC led to the expectation that a similar property would hold also in the full theory. Consequently, instead of the quantum Hamiltonian, a classical one (given by the expectation value of the quantum Hamiltonian operator on a family of semiclassical states) was used \cite{AC14, DL17a}.

Preliminary results in this approach indicated that, when semiclassical states peaked on cosmological data are used, LQG reproduces on the qualitative level the effective dynamics of LQC within the so-called $\mu_o$-scheme \cite{Bojowald:1999tr,Ashtekar:2003hd}. Unfortunately, in LQC, this scheme has proved to be physically inconsistent \cite{Corichi:2008zb}, consequently being replaced by the so-called $\bar\mu$-scheme \cite{APS06c}. It is an open question whether LQG effective dynamics can lead to a physically consistent effective model (e.g., by qualitatively reproducing LQC within the $\bar\mu$-scheme or one of its possible extensions \cite{Engle:2019zfp}).

For technical reasons, the studies in this direction have so far been limited to so-called graph-preserving\footnote
{
In LQG, the space of states consists of cylindrical functions supported on graphs. A graph-preserving operator is an operator which preserves the subspace of cylindrical functions supported on each given graph \cite{AQG1}.
}
Hamiltonians. In the current paper, we investigate whether the commonly known techniques, when applied to these Hamiltonians, can lead to a physically consistent effective model.

The structure of the paper is as follows. In section \ref{s2_mu0} we recall the original conjecture of effective dynamics in the $\mu_o$ scheme and its relation to LQC. In section \ref{s3_tomek} we investigate whether a similar conjecture for the $\bar{\mu}$ scheme can be formulated in the full theory following a proposal from \cite{Alesci} and find the answer in the negative. In section \ref{s4_nogo} we put the problems of finding the $\bar \mu$ scheme in the full theory on a broader ground by presenting explicit no-go statements. Finally, we conclude with possible alternatives in \ref{s5_conclusion}.

Throughout the paper, we work in natural units ($\hbar = G = 1$).
\section{Effective dynamics}
\label{s2_mu0}
Let us start by briefly reviewing the content of effective dynamics in LQC. Classically, in isotropic models the geometry data are contained in a pair of canonical variables: these can be either the triad and connection components \cite{Ashtekar:2003hd}, $p$ and $c$, or the scaled, oriented volume $v \propto p^{3/2}$ and dimensionless $b$ proportional to the Hubble rate\footnote{The most popular convention is: $v=2\pi G\hbar\gamma\sqrt{\Delta}p^{3/2}$ and $b=c\sqrt{\Delta/p}$, where	$\Delta$ is the so-called ``area-gap'' \cite{APS06c}.}. Application of the canonical formalism leads to a constrained system: in order to introduce a meaningful notion of dynamics, one couples the geometry to a convenient set of matter fields (so-called ``internal clocks'') and solves the scalar constraint by group-averaging \cite{Ashtekar:1995zh}. This procedure leads to ``deparametrization on the quantum level'' where the dynamics of the system is generated by a true Hamiltonian, with one clock field playing the role of time. At the deparametrized level, states at a fixed value of the clock become physical states, and we denote their space by $\mathcal{H}_{\rm LQC}$. All relevant geometric operators on $\mathcal{H}_{\rm LQC}$ can be written in terms of two fundamental operators. The choice of these operators is a consequence of the particular regularization scheme: in $\mu_o$-scheme they are $\hat p$ and $\hat{\mathcal N}_1 := \widehat{e^{ic\mu_o}}$ ($\mu_o$ being a positive constant); in the $\bar\mu$-scheme they are $\hat v$ and $\hat{\mathcal N}_2 := \widehat{e^{ib}}$.

A substantial set of cosmological models has been already analyzed within the LQC framework. This includes in particular (but is not restricted to) the models of isotropic universe (the so called Friedman-Lemaitre-Robertson Walker (FLRW) model) of various topologies of constant time slices \cite{Bojowald:2002gz,APS06c,Ashtekar:2006es,Szulc:2006ep,Szulc:2007uk,Bojowald:2008ec}, with various matter content \cite{Bojowald:2002xz,Husain:2011tm,Pawlowski:2014fba} and possibly admitting non-vanishing cosmological constant \cite{Bentivegna:2008bg,Pawlowski:2011zf}, as well as homogeneous anisotropic models (including the so called Bianchi I, II, and IX) \cite{Bojowald:2003md,Chiou:2006qq,MartinBenito:2008wx,Ashtekar:2009vc,Ashtekar:2009um,WilsonEwing:2010rh}. For the models listed above there exists a set of states $\{\psi_{\bf c,p} \in \mathcal{H}_{\rm LQC}\}_{({\bf c,p})\in\mathbb{R}^2}$ (e.g. coherent states peaked about $p={\bf p}$ and $c={\bf c})$ such that, for any observable $O$ polynomial in the fundamental operators, it is ($I\in \{1,2\}$)
\begin{align}\label{peakednessLQC}
\langle\psi_{\bf c,p}, \;& O(\hat{\mathcal{N}}_{I}, \hat v ) \;\psi_{\bf c,p}\rangle_{\rm LQC} = 
\\
& O(e^{i\mu_I(p) c},{p}^{3/2})|_{(c,p) = ({\bf c},{\bf p})}+\mathcal{O}(\vec\epsilon_I)\nonumber
\end{align}
where $\mu_1(p) = \mu_o$ constant and $\mu_2(p) = \sqrt{\Delta/p}$.

Note that $\vec\epsilon_I$ is a vector of the second-order corrections, i.e., relative dispersions and covariances of fundamental operators forming the polymer analogue of the Heisenberg algebra \cite{BS06,BBHKM11}: $\hat p$ or $\hat v$ (for $I = 1$ and $I = 2$ respectively), $(\hat{\mathcal N}_I + \hat{\mathcal N}_I^\dag)/2$ and $(\hat{\mathcal N}_I - \hat{\mathcal N}_I^\dag)/2i$. In the following we consider the states (called ``semiclassical'') for which the remainder $\mathcal O(\vec\epsilon_I)$ is small;\footnote{By the remainder $\mathcal{O}(\vec{\epsilon_I})$ we mean any functions depending on the generalized Hamburger moments, such that it vanishes if the moments are put to zero.
See Appendix \ref{app1} for details.} for simplicity, in the following we drop the symbol $\mathcal{O}(\vec \epsilon_I)$ and use $\approx$ instead of $=$ when an identity holds to zeroth order in $\vec\epsilon_I$.

For certain models admitting massless scalar field (including the flat FLRW universe with non-negative cosmological constant or negative curvature), the semiclassicality property defined above may not be preserved by the dynamics (see for example \cite{Bojowald:2010qm,Kaminski:2019tqo}, also the discussion in \cite{Pawlowski:2011zf,ADLP19}). In these cases the Dirac observables corresponding to $p(t)$ may be ill defined on the physical Hilbert space, thus alternative observables encoding the same information need to be used \cite{Pawlowski:2011zf,ADLP19}. Other choices of matter fields for an internal clock (like dust \cite{Husain:2011tm} or radiation \cite{Pawlowski:2014fba}) are free from this deficiency.

Despite the above problem, probing the quantum dynamics in LQC shows that for many of the models listed above
\begin{align}\label{effectiveLQC}
& \langle \psi_{\bf c,p}, e^{it H(\hat{\mathcal{N}}_I,\hat v)}O(\hat{\mathcal{N}}_I, \hat v) e^{-it H(\hat{\mathcal{N}}_I,\hat v)} \psi_{\bf c,p}\rangle_{\rm LQC} \nonumber
\\
& \approx O(\alpha^t_h[e^{i\mu_I(p) c}],\alpha^t_h[p^{3/2}])|_{(c,p) = ({\bf c},{\bf p})}
\end{align}
where $\alpha^t_h[f]:=\exp(t\{h,.\})(f)$ is the Hamiltonian flow generated by the effective Hamiltonian $h(c,p,\mu_I) := \langle\psi_{c,p}, H(\hat{\mathcal{N}}_I,\hat v) \psi_{c,p}\rangle_{\rm LQC}$ on the phase space coordinatized by $(c,p)$ \cite{Taveras}. Correctness of (\ref{effectiveLQC}) was tested in several models, 
including the models of isotropic universe of various topologies ($K=0,\pm 1$) \cite{APS06c,Ashtekar:2006es,Diener:2014mia,Diener:2014hba}, various values of cosmological constant \cite{Bentivegna:2008bg,Pawlowski:2011zf} and several forms of matter content: dust \cite{Husain:2011tm}, radiation \cite{Pawlowski:2014fba} and massless scalar field, see e.g. \cite{MartinBenito:2009aj}. It was also tested in some homogeneous nonisotropic models -- Bianchi I universe \cite{MartinBenito:2009qu,tp-talk,Diener:2017lde,henderson-tp}. For other models, like the one describing the flat Bianchi I universe with massless scalar field (including the isotropic sector) the result \eqref{effectiveLQC} can be obtained with a minor modification to the present mathematical procedure of building a physical Hilbert space. For other cases (universe of negative curvature or positive cosmological constant) an analogous result holds once the observable $\hat{v}$ is replaced with its 'compactified' analog (see \cite{Pawlowski:2011zf,ADLP19}).
These results have given rise to the {\it effective dynamics conjecture}, namely, that an analogous property also holds for other reduced models (of which dynamics was not tested on the genuine quantum level).\\

Let us now turn towards the full theory. In LQG, given a fixed graph, the fundamental operators are holonomies $\hat h(e)$ of Ashtekar connection along edges $e$ of the graph and fluxes $\hat E(e)$ of the densitized triads across surfaces dual to each link \cite{Thiemann:2007zz, Ashtekar:2004eh}. Given that LQC inherits its structures from LQG, the existing attempts of realizing effective dynamics in LQG rely on a similar framework as the one presented above. So far, all approaches in the literature select for that purpose compact (implicitly embedded in a $3$-torus) cuboidal lattices with $N$ vertices. On the Hilbert space $\mathcal H_N$ of one such graph, one considers a family $\{\Psi^N_{\boldsymbol{\xi},\boldsymbol{\eta}} \in \mathcal H_N\}_{(\boldsymbol{\xi},\boldsymbol{\eta}) \in \mathfrak{su}_2^{3N} \times \mathfrak{su}_2^{3N}}$ of states, that satisfy a semiclassicality property analogous to (\ref{peakednessLQC}), namely
\begin{align}\label{Peakedness}
\langle \Psi^N_{\boldsymbol{\xi},\boldsymbol{\eta}}, O(\hat{h},\hat{E}) \Psi^N_{\boldsymbol{\xi},\boldsymbol{\eta}} \rangle \approx O(e^{\boldsymbol{\xi}},\boldsymbol{\eta})
\end{align}
for any polynomial $O$ in holonomies $\hat h$ and fluxes $\hat E$. Note that in this equation (and in all that follow) the symbol $\approx$ means that the relation holds up to a remainder depending on relative dispersions and covariances of the fundamental operators. Also, abusing the notation, we use $h\equiv \{h(e)\}_e$, $\xi\equiv \{\xi(e)\}_e$ and similar for $E,\eta$ instead of explicitly referring to each edge $e$.\\

Since we are focusing on the isotropic cosmology sector of LQG effective dynamics, we now restrict our attention to subfamilies of such states which are peaked about isotropic cosmological geometries. This means that the peak holonomy and flux labels $(\boldsymbol{\xi},\boldsymbol{\eta})$ can be expressed in terms of the coordinates on the phase space of isotropic cosmology: $\boldsymbol{\xi}(e) = \mu_o {\bf c} \tau$ and $\boldsymbol{\eta}(e) = \mu_o^2 {\bf p} \tau$, where $\tau(e = e_k)=-i\sigma_k/2$ is a generator of $\mathfrak{su}_2$ (which in general depends on the direction  $k$ of $e_k$) and $\mu_o = N^{-1/3}$ is the coordinate length of edge $e$ with respect to a certain fiducial metric.\footnote
{
In the treatment presented in the literature, a specific embedding is chosen, such that the lattice is regular.
}

Upon these choices, preliminary studies performed for example on the states in \cite{DL17b} indicate that 
\begin{enumerate}[(i)]
\item %
for two polynomials $O_1$ and $O_2$ in the fundamental variables 
\begin{align}\label{Commutator}
& \langle \Psi^N_{\boldsymbol{\xi},\boldsymbol{\eta}} ,i [O_1(\hat h, \hat E), O_2(\hat h, \hat E)] \; \Psi^N_{\boldsymbol{\xi},\boldsymbol{\eta}} \rangle \approx\\
& \{O_1(e^{\mu_o c \tau},  \mu_o^2p \tau), O_2(e^{\mu_o c \tau},  \mu_o^2p\tau)\}|_{(c,p) = ({\bf c}, {\bf p})} , \nonumber
\end{align}
\item %
a certain form of effective dynamics (i.e., analogue to (\ref{effectiveLQC})) might hold. (see e.g. \cite{Han:2019vpw,DKL}).
\end{enumerate}
The latter can be captured in the following:
\begin{conjecture}\label{conj}
Consider a semiclassical state $\Psi^N_{\boldsymbol{\xi}, \boldsymbol{\eta}}$ peaked about isotropic geometry data $({\bf c}, {\bf p})$ and Hamiltonian operator $\hat H = H(\hat h, \hat E)$. For any polynomial $O$ in the fundamental variables, the following holds:
\begin{align}\label{evolution}
& \langle \Psi^N_{\boldsymbol{\xi},\boldsymbol{\eta}} ,\; e^{it\hat H} O(\hat h, \hat E) e^{-it\hat H} \; \Psi^N_{\boldsymbol{\xi},\boldsymbol{\eta}} \rangle \approx
\\
& O(\alpha^t_{H_{\mu_o}}[e^{\mu_o c \tau}],\alpha^t_{H_{\mu_o}}[\mu_o^2p \tau])|_{(c,p) = ({\bf c},{\bf p})} \notag
\end{align}
where $\alpha^t_{H_{\mu_o}}[f]:=\exp(t\{H_{\mu_o},.\})(f)$ is the Hamiltonian flow generated by the effective Hamiltonian $H_{\mu_o} := H(e^{\mu_o c \tau},\mu_o^2p\tau)$ on the phase space coordinatized by $(c,p)$.
\end{conjecture}
Several studies appeared in LQG which make (sometimes implicit) use of this conjecture \cite{AC14,DL17a}, concluding that the LQG quantum dynamics of semiclassical states (supported on a single lattice) resembles the $\mu_o$-scheme of LQC. This scheme, however, was shown to lead to physically inconsistent results within LQC (for example, it does not admit a proper infrared regulator removal limit \cite{Corichi:2008zb}). It would therefore be desirable to reproduce in LQG the $\bar\mu$-scheme. In other words, we would like to find a set of semiclassical states in the full theory such that (we omit the explicit symbol for such state)
\begin{align} \label{wishful-thinking}
& \langle e^{it\hat H} O(\hat h, \hat E) e^{-it\hat H} \rangle \approx \\
& O(\alpha^t_{H_{\bar \mu}}[e^{\bar \mu c \tau}],\alpha^t_{H_{\bar \mu}}[\bar \mu^2p] \tau)|_{(c,p) = ({\bf c}, {\bf p})}\notag
\end{align}
where $H_{\bar\mu} := H(e^{\bar\mu c\tau},\bar\mu^2p\tau)$ and $\bar\mu := \bar\mu(p) = \sqrt{\Delta/p}$. In other words, the quantum dynamics of this semiclassical state would be described by the $\bar\mu$-scheme effective Hamiltonian $H_{\bar\mu}$. Such a feature, however, has an unfortunate consequence: from (\ref{wishful-thinking}), by setting $t = 0$ and $O(\hat h, \hat E) = \hat h(e)$, it follows that
\begin{align} \label{weight-hol}
\langle \hat h(e)  \rangle \approx e^{\bar{\mu}({\bf p}){\bf c} \tau(e)}
\end{align}
which means that labels ${\bf c}$ and ${\bf p}$ do not have the meaning of connection and triad coefficients as provided in \cite{Ashtekar:2003hd}.

 Alternatively, if we want to retain the meaning of ${\bf c}$ and ${\bf p}$, equation (\ref{weight-hol}) suggests reinterpreting the multiplication operator $\hat h$ in terms of a new classical object, which we might call a ``weighted holonomy''.\footnote
{
In the context of the full theory, the expression of this weighted holonomy is not specified: we only know that it should reduce to $e^{\bar{\mu}({\bf p}) {\bf c} \tau(e)}$ in the cosmological sector.
}
This is an important departure from standard LQG, that cannot be dismissed easily. For example, one must make sure that $H(h,E)$ remains a regularization of the general relativity (GR) Hamiltonian if $h(e)$ is the weighted holonomy (especially considering the fact that Thiemann identities only work with regular holonomies \cite{QSD}). Nevertheless, let us assume that this issue can be overcome: the system will still be quantized in the usual way, i.e., in terms of ${\rm SU}(2)$ multiplication operators and right-invariant vector fields. Hence, on the quantum level the commutator structure has no knowledge of its former classical origin. This approach will be further discussed in the conclusion.

At the moment we focus on recovering property (\ref{wishful-thinking}) itself. Thus, we now look for possible techniques considered viable to achieve this goal.
\section{A multi-sector strategy}
\label{s3_tomek}
One of the most promising procedures is to consider states with support on a collection of graphs instead of a single one \cite{Alesci}. Since the graph-preserving Hamiltonians (and the standard set of observables) by definition leave the subspaces of states supported on each graph invariant under their action (making each subspace a superselection sector), we can for simplicity call such approach a ``multi-sector strategy'', in opposition to a single-sector one, where just one superselection sector is considered. 

For the class of graph topologies considered in this paper (compact cuboid lattices enumerated by a number of vertices $N$), such states are of the form
\begin{equation}
\hat \rho = \sum_{N=1}^{\infty} c_N(w) |\Psi^N_{\boldsymbol{\xi},\boldsymbol{\eta}} \rangle \langle \Psi^N_{\boldsymbol{\xi},\boldsymbol{\eta}}|
\end{equation}
where we adopted a density matrix notation. Here, $w$ denotes an abstract label which may, in principle, be a function of the phase variables $c,p$ and of the coherent state labels ${\bf c}, {\bf p}$. The hope behind this generalization was based upon the expectation that, given a well-behaved function $F(c,p,\mu_o(N))$ with $\mu_o(N) = N^{-1/3}$, it would be possible to find a family $c_N(w)$ such that
\begin{equation} \label{alesci-obs}
  \sum_{N=1}^{\infty} c_N(w) F({\bf c},{\bf p},\mu_o(N)) \approx F({\bf c},{\bf p},\mu(w)) 
\end{equation}
where $\mu(w)$ (being determined by  the choice of $c_N(w)$) would take a desired form consistent with the $\bar{\mu}$ scheme of LQC. Indeed, for an observable $\hat{O}$ being an operator polynomial in $\hat{h}, \hat E$ one has
\begin{equation}\label{eq:obs-sectors}\begin{split} 
\langle O(\hat h, \hat E) \rangle & := {\rm Tr}[\hat \rho\; O(\hat h,  \hat E)] = \\
& \sum_{N=1}^{\infty}c_N(w) \langle  \Psi^N_{\boldsymbol{\xi},\boldsymbol{\eta}}, O(\hat h, \hat E)\Psi^N_{\boldsymbol{\xi},\boldsymbol{\eta}}\rangle \approx
\\
& \sum_{N=1}^{\infty}c_N(w) O(e^{\mu_o(N) {\bf c} \tau},\mu_o^2(N) {\bf p} \tau)\approx
\\
& O(e^{\mu(w) {\bf c} \tau},\mu^2(w) {\bf p} \tau)
\end{split}\end{equation}
where in the third line we used (\ref{Peakedness}) and in the fourth we used (\ref{alesci-obs}). This shows that one has a significant freedom of affecting the expectation value of $\hat{O}$ by selecting the distribution $c_N(w)$ (e.g., requiring it to be peaked about an appropriate function of $\bf p$).

The first example of applying this strategy discussed in the literature was presented in \cite{Alesci} and relied on a specific postulated choice of $c_N(w)$:
\begin{equation} \label{coefficients}
    c_N(w)=\frac{1}{2^{(\alpha w)^{2/3}}} \binom{(\alpha w)^{3/2}}{N}
\end{equation}
(with $\alpha> 0$). 
This choice led to the desired result $\mu(w) = \sqrt{\Delta/w} =: \bar\mu(w)$ for time-zero expectation values, since then
\begin{equation}
    \langle O(\hat h, \hat E) \rangle \approx O(e^{\bar\mu(w) {\bf c} \tau},\bar\mu^2(w) {\bf p} \tau)
\end{equation}
Upon identifying $w={\bf p}$ and applying this equation to the Hamiltonian operator $\hat O = \hat H$, this expectation value is found to coincide (up to subleading corrections) with the LQC effective Hamiltonian in the $\bar \mu$-scheme.

This is an encouraging result, however what we really need to show is (\ref{wishful-thinking}) whose right hand side, in particular, implies a non-trivial dependence ${\bf p}(t) := \alpha^t_{H_{\bar \mu}}[p]|_{(c,p) = ({\bf c},{\bf p})}$.\footnote%
{
In order to be able to provide a viable description of the observed reality, the model needs to give dynamical predictions which in the low energy limit are converging to those of (the cosmological sector of) classical general relativity. The latter in turn predicts a highly nontrivial time dependence of the values of $c$ and $p$.
}
Therefore, in order for proposals such as (\ref{coefficients}) to yield the $\bar{\mu}$-scheme at arbitrary times, one needs to identify $w={\bf p}(t)$. This in turn implies that the coefficients $c_N(w)$ must have non-trivial time dependence when the evolution is considered. We are now going to show that the quantum evolution described by the left hand side of (\ref{wishful-thinking}) cannot allow for such time-dependence.

Recall that $\hat \rho_t:= e^{-it\hat H}\rho e^{it\hat H}$ and introduce projectors $\hat P_N=\sum_i |e_{N,i}\rangle\langle e_{N,i}|$ onto each graph, so that
\begin{align}
  &\mathbb{I} = \sum_{N} \hat P_N
\end{align}
The unitarity of quantum time evolution requires that, for the coefficients $c_N(t):=c_N(w(t))$, it holds
\begin{align}
  \begin{split}
    & c_N(t)=Tr[\hat \rho_t \hat P_N] \\
    & = \sum_{M,i} c_M(0) \langle \Psi^M_{\boldsymbol{\xi},\boldsymbol{\eta}} | e^{it\hat H} | e_{N,i}\rangle \langle e_{N,i} | e^{-it\hat H} | \Psi^M_{\boldsymbol{\xi},\boldsymbol{\eta}}\rangle\\
    & =c_N(0) \| e^{-it\hat H} \Psi^N_{\boldsymbol{\xi},\boldsymbol{\eta}} \|^2 = c_N(0) \|\Psi^N_{\boldsymbol{\xi},\boldsymbol{\eta}}\|^2 \\
    & = c_N(0)
  \end{split}
\end{align}
This shows that $c_N$ cannot depend on time and hence (\ref{wishful-thinking}) cannot be satisfied by such states $\hat \rho$.

The explicit computation of the expectation value of $\hat O$ on $\hat \rho_t$ gives
\begin{equation}\label{mixedevolution}
\begin{split}
& O(t) := {\rm Tr}[ \hat \rho_t\; O(\hat h, \hat E)]={\rm Tr}[\hat \rho\; e^{it\hat H}O(\hat h, \hat E)e^{-it\hat H}]
\\
& = \sum_{N=1}^{\infty}c_N(w) \langle  \Psi^N_{\boldsymbol{\xi},\boldsymbol{\eta}}, e^{it\hat H} O(\hat h, \hat E)e^{-it\hat H} \Psi^N_{\boldsymbol{\xi},\boldsymbol{\eta}}\rangle
\\
& \approx \sum_{N=1}^{\infty}c_N(w)O\left(\alpha^t_{H_{\mu_o(N)}}[e^{\mu_o(N)c \tau}],\right.
\\
& \hspace{3cm}\left.\alpha^t_{H_{\mu_o(N)}}[\mu_o^2(N)p \tau]\right)|_{(c,p) = ({\bf c}, {\bf p})}
\\
& = O\left(\alpha^t_{H_{\mu(w)}}[e^{\mu(w)c \tau}],\alpha^t_{H_{\mu(w)}}[\mu(w)^2(N)p \tau]\right)|_{(c,p) = ({\bf c}, {\bf p})}
\end{split}
\end{equation}
where in the third line we used (\ref{evolution}) and in the last line we used equation (\ref{alesci-obs}). It could be argued that the choice $w = p$ would lead to the correct result. However, the state $\hat \rho_t$ (and therefore $c_N$ as well) depends only on ${\bf p}$, ${\bf c}$ and $t$. The phase space functions $c,p$ are merely intermediate, auxiliary objects (meaningful only inside each term of the sum in the third line), consequently $w$ cannot be a function on the phase space coordinatized by $(c,p)$: the only option is therefore $\mu({\bf p})$, for which \eqref{mixedevolution} gives
\begin{align} \label{expecatation-t}
& \langle e^{it\hat H} O(\hat h, \hat E) e^{-it\hat H} \rangle = {\rm Tr}[\hat \rho_t\;O(\hat h, \hat E)] \approx
\\
& O(\alpha^t_{H_{\mu({\bf p})}}[e^{\mu({\bf p}) c \tau}], \alpha^t_{H_{\mu({\bf p})}}[\mu({\bf p})^2 p \tau])|_{(c,p) = ({\bf c},{\bf p})}\notag
\end{align}
It is now clear that $\mu({\bf p})$ Poisson-commutes with the functions on which $\alpha^t$ acts, and hence does not contribute to the effective dynamics:
\begin{observation}\label{obs1}
If the quantum dynamics on a single sector (graph) reproduces the $\mu_o$-scheme (Conjecture \ref{conj}), then the quantum dynamics on the multi-sector also reproduces the $\mu_o$-scheme (with a different constant $\mu_o' := \mu({\bf p})$).
\end{observation}
\section{No-Go Statements}
\label{s4_nogo}
The approach discussed so far does not reproduce the $\bar\mu$-scheme, that is, the expectation values of observables $\hat O$ on quantum-evolved states are not consistent with (\ref{wishful-thinking}), but rather reproduce the effective dynamics of the $\mu_o$-scheme. However, to arrive to this conclusions, the use of Conjecture 1 was central. While this conjecture is supported by numerical evidence in the context of symmetry-reduced models, and promising work on providing a proof to it is ongoing \cite{Han:2019vpw,DKL}, one still cannot exclude the possibility that for certain classes of (sharply peaked) states the dynamics may follow different trajectories. However, we will show that, under some weaker assumptions, certain no-go statements concerning the recovery of the $\bar{\mu}$-scheme from the full theory can be made.

For simplicity, in the following we focus on the single-sector pure states. The extension to mixed states can be performed by a procedure similar to that presented in the previous section and, as it was shown there, would not lead to a qualitative change of predictions.

The first no-go statement considers an alternative to the original Conjecture \ref{conj}, in which the semiclassical states peakedness is defined with respect to different phase space coordinates.
\begin{observation}\label{nogo1}
	Let $\Psi^N_{\boldsymbol{\xi},\boldsymbol{\eta}}$ be a semiclassical state with $\boldsymbol{\xi} = \bar{\mu}({\bf p}) {\bf c} \tau =: b({\bf c},{\bf p})$ and $\boldsymbol{\eta} = \bar{\mu}({\bf p})^2{\bf p} \tau =: \tilde{\eta}({\bf p}) \tau$, satisfying property (\ref{Peakedness}). Then, equation
	\begin{align}\label{Commutator-2}
	& \langle \Psi^N_{\boldsymbol{\xi},\boldsymbol{\eta}} ,i [O_1(\hat h, \hat E), O_2(\hat h, \hat E)]  \Psi^N_{\boldsymbol{\xi},\boldsymbol{\eta}} \rangle \approx\\
	& \{O_1(e^{b(c,p)}, \tilde{\eta}(p)\tau), O_2(e^{b(c,p)}, \tilde{\eta}(p)\tau)\}|_{(c,p) = ({\bf c}, {\bf p})} 
	\nonumber
	\end{align}
	cannot be satisfied, which means that the space of labels $(b,\tilde \eta)$ cannot serve as the coordinates of the effective phase space of cosmology.
%	Then, any Hamiltonian $\hat H=H(\hat h, \hat E)$ produces either trivial dynamics for observables $O(\hat h, \hat E)$ or the dynamics differs from the analogue of (\ref{wishful-thinking}) by 0th order terms (i.e., terms that do not vanish once the relative dispersions, covariances and higher order quantum corrections are sent to zero). In other words: 
%	\begin{align}\label{evolution-2}
%	&\langle \Psi^N_{\boldsymbol{\xi},\boldsymbol{\eta}} , e^{it\hat H} O(\hat h, \hat E) e^{-it\hat H} | \Psi^N_{\boldsymbol{\xi},\boldsymbol{\eta}} \rangle \not\approx\\ 
%	&O(\alpha^t_{H_{\bar \mu}}[e^{b(c,p)}], \alpha^t_{H_{\bar \mu}}[\tilde{\eta}(p)\tau])|_{(c,p) = ({\bf c},{\bf p})} \notag
%	\end{align}
%	where $H_{\bar\mu} = H(\exp(b(c,p)),\tilde{\eta}(p)\tau)$.
\end{observation}
The reason why this statement holds is relatively straightforward: recalling that $\bar\mu ({\bf p}) =\sqrt{\Delta/{\bf p}}$, we have $\tilde{\eta}({\bf p}) = \bar{\mu}({\bf p})^2{\bf p} = \Delta$; this, however, means that $\tilde{\eta}(p)=\tilde{\eta}(0)$ is independent of $p$, which makes $\tilde \eta$ unsuitable as a coordinate on the phase space (thus making the coordinate system degenerate). In particular, any Poisson-bracket in (\ref{Commutator-2}) is necessarily zero (and similarly the Hamiltonian flow would preserve $\tilde{\eta}$: $\alpha^t_{H_{\bar\mu}}(\tilde{\eta})=\tilde{\eta}$).

Noting that on a single sector the expectation value of the volume of the spatial manifold is $\langle \hat V [\sigma]\rangle \appropto N$, one may try to implement a multi-sector strategy (such as the one discussed before), constructing a family of states peaked about coordinates $(b,N_o)$. Conceivably, a canonical Poisson structure can be defined on this space, therefore avoiding the problems of Observation \ref{nogo1}. However, due to the non graph-changing nature of the Hamiltonian, the expectation value of the number operator $\sum_N N \hat P_N$ is a constant of motion, and hence ${\bf p} = {\bf N}_o^{2/3}$ would have trivial dynamics, in contradiction with the low energy GR limit.

To summarize: considering states semiclassical in variables $b(c,p)$ and $\eta(p)$ more suitable from the physical point of view, will not lead to any replacement of Conjecture 1 (equation (\ref{evolution})) consistent with the $\bar\mu$-scheme.

At first glance it appears to be possible nonetheless to achieve the $\bar \mu$ scheme by dropping (\ref{Commutator-2}), i.e., by no longer relating $\tilde{\eta}(p)$ on the right hand side with the $\boldsymbol{\eta}$ on which $\Psi$ is peaked. However, we will demonstrate that this cannot be correct in general, using as example a certain regularisation of the Hamiltonian in LQG and the volume operator:\footnote{Work in cosmology is mostly concerned with the volume, however for physical predictions any working conjecture should in principle be extended to Ricci scalar and energy density.}
\begin{observation}\label{nogo2}
	Consider a state obeying (\ref{Peakedness}) with $\boldsymbol{\xi}=\bar\mu({\bf p}) {\bf c} \tau$ and $\boldsymbol{\eta}=\bar \mu^2({\bf p}) {\bf p} \tau = \tilde{\eta}(\boldsymbol{p}) \tau$ and $N=\bar\mu ({\bf p})^{-3/2}$, such that:
	\begin{align}
	&\langle \hat V[\sigma] \rangle \approx N \tilde{\eta}^{\frac{3}{2}} = {\bf p}^{\frac{3}{2}}
	\end{align}
	Then, for a Hamiltonian $\hat H=H(\hat h, \hat E)$ it is
	\begin{align}\label{volume-evolution}
	&\langle e^{it\hat{H}}\hat V[\sigma] e^{-it\hat H} \rangle \not\approx \alpha^t_{H_{\bar \mu}}(p^{\frac{3}{2}})\big|_{(c,p)= ({\bf c},{\bf p})}
	\end{align}
	where $H_{\bar\mu} = H(e^{\sqrt{\Delta/p} \, c\,\tau},\Delta\,\tau)$.\footnote{We refer to the {\it same} function $H$ on the classical phase space, which was used to define the quantum dynamics. Of course, this does not exclude the possibility that (\ref{volume-evolution}) with an "$\approx$" is satisfied for some different effective Hamiltonian $H_{\bar{\mu}}:=H'(e^{\sqrt{\Delta/p}c\tau},\Delta \tau)$ on the right hand side.}
\end{observation}
In other words, given an isotropic state initially peaked in volume at ${\bf p}^{3/2}$ and assuming its peak follows some effective trajectory under quantum dynamics for some Hamiltonian (which is a function of ${\rm SU}(2)$ multiplication operators and right-invariant vector fields), such trajectory will not be the one which is generated by replacing the operators with the respective classical expressions of isotropic holonomies and fluxes in the $\bar \mu$-scheme.

If both sides of (\ref{volume-evolution}) were equal for all $t$, then the expansion in $t$ of (\ref{volume-evolution}) must coincide order by order:
\begin{align}\label{CommutatorProblemGeneral}
\langle [\hat H, \hat V[\sigma]]_{(n)} \rangle \approx \{H_{\bar\mu},p^{\frac{3}{2}}\}_{(n)} |_{(c,p)=({\bf c},{\bf p})}
\end{align}
where $\{A,B\}_{(n)}$ is defined inductively by $\{A,B\}_{(n+1)} = \{A,\{A,B\}_{(n)}\}$ and $\{A,B\}_{(1)} = \{A,B\}$, and $[\hat A, \hat B]_{(n)}$ is defined analogously. In particular, we must have
\begin{align}\label{CommutatorProblem}
&\langle [\hat H, [\hat H, \hat V[\sigma]]] \rangle \approx \{H_{\bar\mu} \{H_{\bar\mu},p^{\frac{3}{2}}\}\} |_{(c,p)=({\bf c},{\bf p})}
\end{align}
To understand better the consequences of this equation, we consider a particular (non-physical) example: the Euclidean Hamiltonian operator $\hat H$ acting on a cubic lattice as proposed by Giesel and Thiemann \cite{AQG1}. This operator has the property that
\begin{align}
H(e^{\sqrt{\Delta/p} \, c\,\tau}, \Delta\,\tau) = \sin^2(\sqrt{\Delta/p} \, c) p^{3/2}/\Delta
\end{align}
so that $H_{\bar\mu}$ is indeed the $\bar\mu$-scheme effective Hamiltonian of LQC. It is then easy to check that the right hand side of (\ref{CommutatorProblem}) is
\begin{align} \label{example1}
& \{\sin^2\big(\sqrt{\frac{\Delta}{p}} c\big) \frac{p^{\frac{3}{2}}}{\Delta}, \{\sin^2\big(\sqrt{\frac{\Delta}{p}} c\big) \frac{p^{\frac{3}{2}}}{\Delta},p^{\frac{3}{2}}\}\}|_{(c,p)=({\bf c},{\bf p})} \notag
\\
& = \frac{\kappa^2\beta^2}{8} \frac{{\bf p}^{\frac{3}{2}}}{\Delta} \sin^2\big(\sqrt{\frac{\Delta}{\bf p}} {\bf c}\big)
\end{align}
On the other hand, the left hand side -- that is, the double commutator between operators in the full theory -- can be computed explicitly (see \cite{LR19} for details). The evaluation gives
\begin{align} \label{example2}
&\langle [\hat H, [\hat H, \hat V[\sigma]]] \rangle
\\
& \approx \dfrac{\kappa^2\beta^2}{8} \dfrac{p^{\frac{1}{2}}}{\mu^2} \sin^2(\mu c) \dfrac{2 + \cos(2\mu c)}{3} \bigg|_{(c,p,\mu)=({\bf c},{\bf p},\sqrt{\frac{\Delta}{\bf p}})} \notag
\end{align}
The mismatch between the two sides of the equation shows that (\ref{CommutatorProblem}) cannot hold.

To analyze the problem in full generality, it is convenient to introduce two maps from operators to phase space functions:
\begin{align}
\omega : O(\hat h, \hat E) \mapsto O(e^{\mu c \tau}, \mu^2 p \tau)
\end{align}
where $\mu$ is considered as a parameter unrelated to phase space coordinates, and 
\begin{align}
\tilde \omega : O(\hat h, \hat E) \mapsto \omega(O(\hat h, \hat E))\big|_{\mu = \sqrt{\Delta/p}}
\end{align}
Up to second-order corrections, these maps associate to a given operator $\hat O$ the expectation value of $\hat O$ on semiclassical states defined in Conjecture \ref{conj} and Observation \ref{nogo2} respectively. Notice that the only difference between the two maps is the identification of $\mu$ with the phase space function $\sqrt{\Delta/p}$ in $\tilde\omega$ (after evaluating $\omega$).

In terms of these maps, equation (\ref{CommutatorProblemGeneral}) takes the following form:
\begin{align} \label{commutator-wish}
\tilde\omega([\hat H, \hat V]_{(n)}) \approx \{\tilde\omega(\hat H), \tilde\omega(\hat V)\}_{(n)}
\end{align}
To verify whether this can be satisfied, we first observe that, due to equation (\ref{Commutator}), the following equality holds:
\begin{align} \label{commutator-mu0}
\omega([\hat H, \hat V]_{(n)}) \approx \{\omega(\hat H), \omega(\hat V)\}_{(n)}
\end{align}
Writing $\tilde\omega$ in terms of $\omega$ and making use of (\ref{commutator-mu0}), the left hand side and right hand side of equation (\ref{commutator-wish}) read respectively
\begin{align}
\tilde \omega([\hat H, \hat V]_{(n)}) & = \omega([\hat H, \hat V]_{(n)})|_{\mu = \sqrt{\Delta/p}} \approx \notag
\\
& \approx \{\omega(\hat H), \omega(\hat V)\}_{(n)}|_{\mu = \sqrt{\Delta/p}}
\end{align}
and
\begin{align}
\{\tilde\omega(\hat H), \tilde\omega(\hat V)\}_{(n)} = \{\omega(\hat H)|_{\mu = \sqrt{\Delta/p}}, \omega(\hat V)|_{\mu = \sqrt{\Delta/p}}\}_{(n)}
\end{align}
These two quantities cannot be equal for all $n$ as long as $\tilde\omega(\hat H)$ is a non-trivial anaytical function of $c$ due to the fact that, with $\mu = \sqrt{\Delta/p}$ being a nontrivial phase space function, for generic $A$ and $B$ we have
\begin{align}
\exists n \in \mathbb N : \{A, B\}_{(n)}|_{\mu = \sqrt{\frac{\Delta}{p}}} \neq \{A|_{\mu = \sqrt{\frac{\Delta}{p}}}, B|_{\mu = \sqrt{\frac{\Delta}{p}}}\}_{(n)}
\end{align}
We therefore conclude that (\ref{commutator-wish}) does not hold, which explains the disagreement between (\ref{example1}) and (\ref{example2}) in the example, and proves Observation \ref{nogo2}.
\section{Conclusion}
\label{s5_conclusion}
In this paper, we investigated whether a physically consistent effective dynamics of cosmological semiclassical states (such has the $\bar\mu$-effective dynamics in LQC) can be obtained from quantum dynamics in full LQG using currently available tools. In particular, we focused on graph-preserving Hamiltonians. Independent studies \cite{DKL} indicate that, for such Hamiltonians, the dynamics on a single superselection sector (i.e., for states supported on a single graph) reproduces the $\mu_o$-effective dynamics up to second-order corrections. This observation was captured in Conjecture \ref{conj}.

Since this outcome is not physically favored, a proposal has appeared \cite{Alesci} to circumvent this problem by considering mixed states defined on ensambles of superselection sectors (i.e., graphs). For such method, we have shown that the requirement of unitarity of quantum evolution forces the dynamics of the mixed state to have the same qualitative features of the single-sector one. In other words, starting from single-sector components obeying Conjecture \ref{conj}, one finds that the mixed state also follows $\mu_o$-effective dynamics (possibly with a different constant $\mu_o'$). This result is summarized in Observation \ref{obs1}.

Following the no-go result of Observation \ref{obs1}, a different route was considered. We studied a different family of semiclassical states, whose peakedness is defined with respect to a different set of phase space coordinates, resembling those of improved dynamics in LQC \cite{APS06a,APS06c} (while keeping the Poisson algebra and the regularization of the Hamiltonian unchanged). We were able to show that the attempt, to identify expectation values of commutators of quantum observables with Poisson brackets of the classical counterparts of these observables expressed as functions of the new coordinates, led to trivial evolution of flux-dependent observables (such as the volume), which is also physically inconsistent. This fact is expressed in Observation \ref{nogo1}. In Observation \ref{nogo2}, it is moreover found that the commutator algebra of the fundamental operators is not consistent with the reduced Poisson structure stemming from the weighted holonomies of the $\bar \mu$ scheme.  Thus, in general the evolution for the volume differs in both descriptions. However, we want emphasize that although the procedure from Observation \ref{nogo2} does not reproduce the $\bar \mu$ scheme, this does not invalidate the possibility that the full theory produces some other effective model which is physically consistent and reproduces GR at low energies.

The methods discussed above cover all the approaches in the literature to graph-preserving Hamiltonians. Since we have shown that none of them leads to consistent physical dynamics, a qualitatively new approach is required. The possibilities include:
\begin{enumerate}[(i)]
	\item Defining a meaningful ``continuum limit'' $\mu \to 0$. Such an approach is expected to lead to classical dynamics in the leading order, while quantum effects would sit in the higher-order corrections.
	\item Considering a graph-changing Hamiltonian. There are several such proposals in the literature, but they all rely on the existence of some ``non-changing core'' to which certain degenerate \cite{QSD} or ultra-local \cite{Assanioussi:2015gka} structures are added. Therefore, these graph-changing Hamiltonians have a problem common with the graph-preserving approach\footnote{This problem has been already noticed in the literature  and in particular has been motivation for ``lattice refinement" \cite{Bojowald:2007ra,Nelson:2007wj}.}: it is not clear whether these solutions are viable from the point of view of describing an expanding universe, since the structure associated with a single node generating non-trivial volume would have to describe a large region of the universe. Therefore, if this route is to be followed, one might need a new proposal for a graph-changing Hamiltonian.
	\item Starting with a new symplectic structure at the classical level, thus applying the quantization procedure to a new algebra of variables. For example, one could replace the holonomy-flux algebra with the algebra of ``weighted holonomies'' and their canonical conjugated momenta, generalizing to the full theory what was done in LQC improved dynamics (see e.g. \cite{Bodendorfer:2014vea,Bodendorfer:2015hwl} for first steps in this direction for reduced models). This, in particular, requires a new regularization of the classical Hamiltonian, in a context where Thiemann identities might not be valid.
	\item Using renormalisation techniques to find a cylindrical consistent choice of graph-preserving Hamiltonians. Those could be used to construct a continuum quantum field theory via inductive limit methods (see e.g. \cite{LLT1}, or in the context of spinfoam formulation to LQG \cite{Bahr:2009qc,Bahr:2017klw}). In this sense, the fixed graphs correspond only to observing the full theory with some coarseness scale $\mu$, while its dynamics is to be computed in the continuum.
\end{enumerate}
These approaches are currently being investigated by several groups. Furthermore, the list above is not exhaustive and there may well exist other 
approaches circumventing the no-go statements (observations) made in this article. Thus, although we have shown a certain popular set of approaches 
to not have a chance to work, there are still other prospects of constructing a framework which will recover a physically consistent scheme for LQG effective dynamics.

\begin{acknowledgments}
	This work is supported by NSF grant PHY-1454832 and the Polish Narodowe Centrum Nauki (NCN) grant 2012/05/E/ST2/03308.
\end{acknowledgments}

\appendix

\section{Effective description of semiclassical states} \label{app1}

Consider a simple quantum mechanical system for which a pair of observables $\hat{x}, \hat{p}$ forms a Heisenberg algebra 
\begin{equation}
	[\hat{x},\hat{p}] = i\hbar \mathbb{I} .
\end{equation}
For a sufficiently rich class of states (which we will define more precisely later) their physical properties can be encoded in the 
set of classical quantities known as generalized Hamburger moments
\begin{align}\label{eq:ham-def}
	&G^{mn} := \text{``}\langle : (\hat{x}-\langle \hat{x} \rangle \mathbb{I})^m (\hat{p}-\langle \hat{p} \rangle \mathbb{I})^n : \rangle \text{''} \\
	&= \sum_{k,l=0}^{m,n} (-1)^{(m+n)-(k+l)} \binom{m}{k}\binom{n}{l} \langle: \hat{x}^k \hat{p}^l : \rangle \langle\hat{x}\rangle^{m-k} \langle\hat{p}\rangle^{n-l} ,\notag
\end{align}
where $:\cdot :$ is a symmetric (usually Weyl) ordering. This decomposition has been known in quantum optics for more than half a century and was reintroduced in context of quantum cosmology in \cite{Bojowald:2005cw}. Remarkably, the countable set of $G^{mn}$ forms a Poisson algebra of complicated but 
known structure. All the observables, which can be written as functions of fundamental operators $\hat x, \hat p$ can be expressed by $G^{mn}$ via an analog of Taylor expansion
\begin{align}\label{eq:ham-exp}
	\langle O(\hat{x},\hat{p}) \rangle 
	&= \text{``} \langle O(\langle\hat{x}\rangle\id + (\hat{x}-\langle\hat{x}\rangle\id),\langle\hat{p}\rangle\id + (\hat{p}-\langle\hat{p}\rangle\id)) \rangle \text{''} \notag\\
	&= \sum_{k,l=0}^{\infty} \frac{1}{k!l!}
		\left.\frac{\partial^{k+l}O}{\partial^k x \partial^l p}\right|_{x=\langle\hat{x}\rangle,p=\langle\hat{p}\rangle} 
		G^{kl} .
\end{align}
Applying this decomposition to the Hamiltonian allows one to write it as a series in $(x=\langle\hat{x}\rangle,p=\langle\hat{p}\rangle, G^{mn})$. Known Poisson structure of the central moments algebra permits one then to find the full (countable) set of equations of motion for $(x,p,G^{mn})$, effectively determining the quantum evolution. In particular, the equations of motion for $(x,p)$ will get contributions in the form of functions of $(G^{mn})$. These terms are the quantum corrections (of the order $m+n$) to the classical trajectories.

This (countable) set can now be truncated at a finite order $m+n$. Provided that the higher order terms in the Hamiltonian as well as the set of moments 
$G^{mn}$ representing the state decay sufficiently fast with the order $m+n$ the resulting truncated system will provide a good approximation of the actual quantum evolution, of which accuracy can be controlled by a truncation order. For that purpose, one usually restricts the studies to the set of states satisfying the inequalities $\forall j,k\in\mathbb{Z}^+\ |G^{m+j,n+k}| \ll \hbar^{j+k} |G^{m,n}|$ providing a stronger notion of {\it semiclassicality}. For many systems the set of such states is sufficiently large to allow for extracting meaningful physical information.

Such description can be generalized in two ways.
First, for the systems featuring classical phase space of higher dimension it generalizes in a straightforward way: the moments $G$ simply become multi-index objects $G^{k_1,\ldots,k_N}$, where $N$ is the classical phase space dimension. 
Second, the formalism can be generalized to quantum representations in which the algebra of fundamental operators has different structure than the Heisenberg one. In particular, in case of the polymer quantization (see for example \cite{G.:2013lia,Corichi:2007tf}) of the system we have a pair of operators: momentum $\hat{p}$ and a boost $\hat{U}_{\lambda} := \widehat{\exp(i\lambda x)}$ with commutator $[\hat{p},\hat{U}_{\lambda}] = -\lambda\hbar \hat{U}_{\lambda}$. One can then introduce a triple of (classical effective) variables as expectation values 
$p:=\langle\hat{p}\rangle, c:=\langle(\hat{U}_{\lambda}+\hat{U}_{\lambda}^{-1})/2 \rangle, s:=\langle(\hat{U}_{\lambda}-\hat{U}_{\lambda}^{-1})/(2i)$ 
and subsequently define the central moments $G^{ijk}$ analogously to \eqref{eq:ham-def}. Subsequently, the observables and the Hamiltonian can be expressed 
as series in the variables $(p,c,s,G^{ijk})$ via expansions analogous to \eqref{eq:ham-exp} and the resulting system of equations of motion can again be truncated. The Poisson algebra structure of $G^{ijk}$ is more complicated, but can be algorithmized and the set of equations of motion truncated at the arbitrary order can be found \cite{brizuela-tp}.

\twocolumngrid
\bibliography{References}{}

\end{document}